\documentstyle[12pt]{article}

\def \ee {e^+ e^-}
\def \mm {\mu^+\mu^-}
\def \pp {p \overline{p}}

\def \L  {{\cal L}}
\def \MT{{\rm M_{top}}}
\def \MH{{\rm M_{Higgs}}}
\def \mmu{{\rm M_{\mu}}}
\def \me{{\rm M_e}}

\def \cmsqs { {\rm cm^{-2} sec^{-1}} }
\def \simleq{\stackrel{<}{\sim}}

\def \numu{\nu_\mu}
\def \numubar{\overline{\nu}_\mu}
\def \nue{\nu_{\rm e}}
\def \nuebar{\overline{\nu}_{\rm e}}

\title{The Physics Opportunities and Technical Challenges of a Muon Collider
\footnote{Paper submitted to Columbia University of New York in partial
fulfillment of the requirements for a Doctorate of Philosophy.}  }
\author{Bruce J.\@ King}
\date{January 24, 1994}
\begin{document}

\maketitle

\vspace{4 cm}

\setlength{\baselineskip}{2\baselineskip}

{\bf Abstract}

  We describe the physics oppportunities and technical
challenges of a muon collider as a tool for exploring
high energy physics phenomena.

\pagebreak

\tableofcontents

\pagebreak

\section{Introduction}

   The continued success of the standard model (SM) of
elementary particle physics has gradually but fundamentally
altered the the character of experimental high energy physics
in the past decade or so. Ever more precise, expensive
and time-consuming experiments continue to agree with
the predictions of the SM, and the only really good
chance for new discoveries appears to be by searching
at energies higher than previously attained (in the
TeV energy range). 

    The high energy frontier also has its problems,
as emphasized by the cancellation of the SSC accelerator.
Colliding beam facilities tend to be very large, technically
challenging and expensive.

   The SSC and the proposed Large Hadron Collider (LHC) at
CERN were designed to collide protons. Proton collisions
have two main drawbacks:
\begin{itemize}
   \item     Protons are complex
        composite particles. The
        hard scattering interactions that could produce new
        high mass particles actually occur between the quark
        and gluon constituents of the proton, and each constituent
        particle carries only a fraction of the proton
        momentum. This lowers the actual collision energy
        and means that interactions occur at a range of
        center of mass (CoM) energies and rest frames.
        The mass reach of hadron colliders for discovering
        new particles is diluted by this, by a factor of
        roughly 10 to 20.

   \item   The strongly interacting protons produce
        enormous numbers of uninteresting background
        particles from soft collisions. This tends
        to obscure the rare interesting processes
        and causes serious radiation and event
        triggering problems for the particle detectors.
\end{itemize}

   The problems of hadron colliders are avoided by colliding
electrons (and positrons). However, electrons have severe
problems with synchrotron radiation which are specifically
related to their light mass ($\me = 0.511$ MeV):
\begin{itemize}
   \item  The energy loss per revolution from synchrotron radiation
for a charged particle in a circular cyclotron accelerator of
radius R is given by
\begin{equation}
{\rm \Delta E(MeV) = 8. 85 \times 10^{-2}\frac{[E(GeV)]^4}{R(meters)} }
\end{equation}
This loss must be compensated for by using radio-frequency cavities
to accelerate the beam. This quickly becomes prohibitive as the
electron energy is increased. The most powerful cyclotron
accelerator for electrons will probably be the LEP-II accelerator
at the CERN laboratory in Switzerland, which will
come on-line in the next few years. The 27 kilometer ring
will provide $\ee$ collisions at CoM energies of 170 GeV.
The only practical way of colliding electrons at energies higher
than this is using single-pass collisions from pairs of opposed
linear accelerators.

   \item
   Even linear electron colliders have the serious problem
of ``beamstrahlung'' at the collision point. In future planned
$\ee$ colliders the magnetic fields generated from the intersection
of high density electron and positron beams will reach thousands
of Teslas, inducing the particles to emit intense synchrotron
radiation. This lowers and spreads out the CoM energies of the
collisions, and also creates a serious background of photons in
the detector.
In addition, the photons can interact with either individual
electrons or the macroscopic electromagnetic field of the
oncoming beam to produce low energy electron pairs, which
also form an experimental background. Pair production
becomes a prohibitive background when the critical
synchrotron radiation energy of the magnetic fields
(equation 14.85 of Jackson\cite{Jackson})
approaches the electron beam energy.

\end{itemize}
    The above problems and the multi-billion dollar expense
of proposed $\ee$ and proton colliders have provoked
a pessimism in the high energy physics community about
the experimental future of the field. Nevertheless,
the importance of further experimental progress to
the advancement of the field cannot be overstated.
To quote Harvard theorist Sidney R. Coleman\cite{Sci. Am.}
``Experiment is the source of imagination. All the
philosophers in the world thinking for thousands of
years couldn't come up with quantum mechanics''.
This impasse underlines the importance of novel
accelerator technologies. In the opinion
of well known experimental physicist Samuel C.
Ting\cite{Sci. Am.}
``We need revolutionary ideas in accelerator design
more than we need theory. Most universities do not
have an accelerator course. Without such a course,
and an infusion of new ideas, the field will die.''

   One idea that shows promise is to avoid the
synchrotron radiation problems of electrons by
using muons instead. These ``fat electrons''
have 200 times the mass of electrons ($\mmu = 105.66$ MeV,
c.f. with 0.511 MeV for electrons) and, in keeping
with the idea of lepton universality, have otherwise
nearly identical physics properties. They can be
produced copiously by impinging proton beams on
a target to produce pions and then letting the
pions decay to muons. The one very serious drawback
of muons is that they are unstable, decaying
with a rest-frame lifetime of 2.2 $\mu {\rm s}$ into electrons
and neutrinos:
\begin{equation}
\mu ^- \rightarrow e ^- + \overline{\nu _e} + \nu _{\mu}.
\end{equation}
This fact means that muon colliders must do
everything very fast. The muons must be collected,
``cooled'' into small dense bunches, accelerated
and collided before a significant fraction of them
decay.


\section{Physics Opportunities at the High Energy Frontier}

    The top quark and the Higgs boson are the two undiscovered
elementary particles required to complete the original (and
simplest) version of the SM -- sometimes called the Minimal
Standard Model (MSM). Experiments have set lower limits on
the masses of the top quark and the Higgs particle of
$\MT \simeq 130$ GeV\cite{Mtop direct search} and
$\MH = 48$
GeV\cite{PDG}, respectively, while the
consistency of the MSM requires $\MT$ to be below
about 250 GeV and $\MH$ to be below $\sim 1$ TeV.
This means that a muon collider could be used to
discover and/or study the properties of either of
these.

   The Fermilab Tevatron $\pp$ collider, operating at
either 900 GeV or 1 TeV, appears to have a reasonable
chance of discovering the top quark in the next few 
years, and it will almost certainly be discovered
if and when the LHC starts taking data around the
turn of the century. However, hadron colliders will
probably only to be able to determine $\MT$ to within
about 5 GeV. The cleaner experimental conditions in
lepton colliders could improve this to better than
1 GeV, and provide better tests of QCD predictions
for top quark decays.

    The Higgs boson is a much more difficult
experimental target because of its low production
cross section. The dominant production modes for
lepton colliders are shown in figures \ref{fig:H lepton}a--d and
the production modes for hadron colliders are
shown in figures \ref{fig:H hadron}a and \ref{fig:H hadron}b.
\begin{figure}
   \vspace{20 cm}
   \caption
{The dominant Higgs production mechanisms for lepton colliders.}
   \label{fig:H lepton}
\end{figure}

\begin{figure}
   \vspace{20 cm}
   \caption
{The dominant Higgs production mechanisms for hadron colliders.}
   \label{fig:H hadron}
\end{figure}

\begin{figure}
   \vspace{10 cm}
   \caption
{Higgs production cross sections for lepton colliders.}
   \label{fig:H xsec}
\end{figure}

The cross section contributions
at lepton colliders from figures~\ref{fig:H lepton}a
and~\ref{fig:H lepton}b are shown
in figure \ref{fig:H xsec}. Note that the higher order process
of \ref{fig:H lepton}b actually rises with increasing CoM energy,
and this is the main Higgs production mechanism
for TeV scale lepton colliders. The cross section
for figure \ref{fig:H lepton}c is smaller than
\ref{fig:H lepton}b because of the
smaller NC coupling and $M_Z > M_W$, and so
it hasn't been considered seriously in the
lepton collider studies I have seen.
(I am not sure
how much smaller -- it is reduced by a factor of
about seven at the HERA ep collider
and I would guess a similar or smaller reduction
at a higher energy lepton collider.) However,
it appears to give a much cleaner signature for
the Higgs particle than the corresponding
W -fusion process because $\MH$ can be
reconstructed from the outgoing leptons and
the known beam energies.
Figure \ref{fig:H lepton}d is enhanced for $\mm$ colliders
relative to $\ee$ colliders by a factor of
$(\mmu / \me )^2 \simeq 40,000$. It makes an
insignificant contribution for electron colliders
but for $\mm$ colliders and $\MH \simleq 200$ GeV
there is a significant Higgs production resonance
at $E_{CM} = \MH$. Once the Higgs has been discovered
a ``Higgs factory'' muon collider could be built
to sit on this resonance.

   The Higgs decays preferentially
to the heaviest particle--antiparticle pair lighter
than $\MH$. At the lighter end of the expected
mass range for $\MH$ the decay to $b \overline{b}$
pairs is favored, while heavier Higgs can decay to
$t \overline{t}$ or W and Z bosons. Hadron colliders
have such enormous background problems for most of
these decays that the Higgs must be searched for in
less common decay modes.

   Another topic in the MSM that lepton colliders
will be particularly useful for studying is the
triple and quartic gauge boson couplings:
$WW\gamma$, $WWZ$, $WWWW$, $WWZZ$, $WW \gamma \gamma$
and $WWZ \gamma$. The anticipated observation of these
couplings at LEP-II  will provide the first experimental
verification of the non-abelian nature of the
standard model, and they can be studied with
greater precision at higher energy lepton
colliders.

   The MSM is known to be only a good phenomenological
theory that becomes inconsistent at experimentally
inaccessible energy scales. The verification of the
MSM at the next generation of colliders is only the
most conservative scenario, and many physicists think
that there is a good chance that exotic new processes
will be revealed. This might take the form of extra
Higgs particles, missing energy from the new particles
predicted in various ``supersymmetric'' theories, or
something even more unexpected. These exciting
possibilities provide some of the main motivation
for building new accelerators.

%
%
%
%
%
%
\section{Luminosity, and Ionization Cooling of Muons}

   The production of high mass particles is expected to
be a very rare process, requiring enormous collision
rates -- this is motivated by the observation that
point-like cross sections fall
as the inverse square of the center of momentum (CoM)
energy. For example, the production of $\ee$ pairs
in muon collisions is given by
\begin{equation}
\sigma(\mm \rightarrow \ee) \equiv
1 R=\frac{4 \pi \alpha^3}{3s} = \frac{87\:fbarn}
                                     {E_{CM}^2\: (TeV^2)}.
         \label{eq:lum}
\end{equation}

The number of
events produced at an accelerator is given by the
product of the cross section for that process, $\sigma$, and
the luminosity of the accelerator, $\L$, integrated
over its running time
\begin{equation}
{\rm number\; of\; events\;} = \sigma  \int \L dt.
              \label{eq:siglum}
\end{equation}

   Design luminosities for the next generation of
planned accelerations are typically
$\L = 10^{33}-10^{34} \cmsqs$. For a
canonical year of $10^7$ seconds this corresponds
to an integrated luminosity of $\sigma  \int \L dt = 10-100$
inverse $fbarn$. (So equation~\ref{eq:lum} predicts
that a muon collider with 1 TeV CoM energy and $\L = 10^{34} \cmsqs$ would
produce around 10,000 electron pairs in a year's running.)

   The luminosity of an accelerator is given by
\begin{equation}
    \L = \frac{N^2 f}{A},
\end{equation}
where $N$ is the number of $\mu^+$ or $\mu^-$ in a bunch
(assumed equal), $f$ is the frequency of collisions and
$A$ is the (effective) cross-sectional area of the beams
at the collision point. The primary goal of accelerator
design is deliver as large an $\L$ as possible at the
specified energy.

   The cross-sectional area, $A$, is minimized by designing
a magnet lattice to focus strongly at the collision point
and by minimizing the phase space volume of the particle bunches
so that they will come to a good focus at the collision
point. The phase space volume, $PS$, of the beam can be written as
a 6-dimensional product of the beam spread in coordinate
and momentum space
\begin{equation}
PS = \Delta x\, \Delta p_x\: \Delta y\, \Delta p_y\: \Delta z\, \Delta p_z.
\end{equation}

    The $PS$ of the particle bunch is conserved in any interactions
with macroscopic external electromagnetic fields, including the
time-dependent fields applied during the acceleration and storage
of the bunch in the accelerator.
The product of the momentum spread and the spatial spread in
each dimension is usually also separately conserved (with a few
caveats), but momentum spread is easily traded for spatial
spread by focusing or defocusing the bunch.
However, $PS$ does tend to increase due to the
following effects
\begin{enumerate}
   \item  The bunch tends to be pushed apart by its own
charge -- the ``space-charge'' effect. This tendency must
be opposed by longitudinal and transverse focusing in the
accelerator.
   \item  Disruptions of the bunches can induced by (e.g.)
interaction of the beam charge with accelerator elements
(particularly r.f. cavities). While in principle this
may not increase the true phase space volume the practical
effect is to cause ``filamentation'' of the bunch so that
it acts as though it is occupying a larger phase space volume.
\end{enumerate}

    Since producing muons from pion decays gives very
large values of $PS$ it is necessary to cool the muons
considerably before acceleration.

    Muons can be cooled by a very simple method known
as ionization cooling. The concept is illustrated
in figure~\ref{fig:ion cool}a. A bunch of muons
is passed through a slab of material to reduce
the muon energies. This reduces the transverse
momentum spread by a factor equal to the fractional
energy loss. The momentum in the direction of the
beam is also reduced, but this can then be restored
by accelerating the bunch in r.f. cavities. The
net effect is that the bunch ends up with the
same energy but a lower transverse momentum spread.
A variation is shown in figure~\ref{fig:ion cool}b.
A wedge of matter is placed in a dispersive region
of the magnet lattice where the high energy muons
are displaced from lower
energy muons. The higher energy muons pass through
more material than the lower energy ones and lose
more energy. The original mean energy is then
restored with an r.f. cavity, and this time the
longitudinal momentum spread of the beam has
been reduced.

\begin{figure}
   \vspace{20 cm}
   \caption
{Ionization cooling of muons.}
   \label{fig:ion cool}
\end{figure}

    This cooling mechanism is unique to muons.
Electrons and hadrons such as protons would interact
in the cooling material, and the only other
heavy lepton -- the tau -- decays far too quickly
for cooling or acceleration.

    There are two heating mechanisms that compete
with the cooling process
\begin{itemize}
    \item  The transverse momentum spread of the beam
is increased by multiple coulomb scattering (MCS)
\begin{equation}
      \frac{d (\Delta p_{x,y})^2}{dz} = \frac{1}{L_R}(13.6\: MeV/c)^2,
\end{equation}
where $L_R$ is the radiation length of the material.

    \item  The longitudinal momentum spread is increased
by energy straggling
\begin{equation}
      \frac{d (\Delta p_z)^2}{dz} = \frac{dE}{dz}I,
\end{equation}
where $I$ is the mean energy exchange ($\sim 12{\rm Z}$ eV), the
additional energy losses from hard single scatters have
been neglected and the approximation $p_z \simeq E$ is used.
\end{itemize}

    Cooling is optimized by
\begin{enumerate}
    \item Using a low Z material such as beryllium to
maximize the energy loss per radiation length and reduce
the energy straggling. (Beryllium has an energy loss of
105 MeV per radiation length, compared with only 7.2 MeV
for lead.)
    \item Focusing the muons into a tight bunch at the
material to blow up the longitudinal and transverse
momentum spreads to large values which can be effectively
reduced by cooling.
    \item Using low energy beams so that the fractional
energy loss per radiation length is maximized. The energy
cannot be below about 0.3 GeV because below this the muons
are no longer relativistic minimum-ionizing particles and
the energy spread of the bunch increases quickly when
passed through material.
\end{enumerate}

    An interesting idea that unfortunately probably
won't work is to use crystals to cool the beam even
further. Certain axes of crystals tend to channel
charged particles and hold them while they lose
energy  -- giving cooling without MCS.
Large, high quality crytals of silicon, germanium
and tungsten have been grown and used for extensive
studies of particle channeling, and bent crystals have
been used to steer particle beams. Unfortunately,
the solid angle for capturing particles is very
small ($\sim$milliradians at 50 MeV, falling
as $1/\sqrt{E}$~cite{Chen crystal})and the
particles dechannel over characteristic lengths
of centimeters at 10 GeV, rising in proportion
to the beam energy\cite{Carrigan}. This appears
to be too small by about two orders of magnitude
for net cooling.

   Beam cooling at a muon accelerator would be
expected to consist of some tens of slabs of beryllium
or some other low Z material inside a lattice of
magnets and accelerating structures to transport
the beam and manipulate its distribution in
phase space.

\section{Conceptual Design of a Muon Collider}

   The idea of muon storage rings has probably been
around since the 1960's or earlier, and muon colliders
have been seriously discussed at least as early as
1980\cite{Krinsky}. A conceptual design of a muon
collider is shown in
figure~\ref{fig:Neuffer acc}~\cite{Neuffer}.
This section discusses each of the components of
the accelerator.
\begin{figure}
   \vspace{20 cm}
   \caption
{Conceptual design of a muon collider.}
   \label{fig:Neuffer acc}
\end{figure}

    The requirement of colliding bunches containing
$10^{11}-10^{12}$ muons means that
the hadron accelerator must deliver $10^{13}-10^{14}$
protons into the target at a rate of 10 Hz or higher.
This is more than any existing accelerator, but this
technology has been studied in detail for the planned
meson factories KAON and PILAC. The KAON design calls
for bunches of $6 \cdot 10^{13}$ 30 GeV protons at a
rate of 10 Hz.

    Possible modifications to the KAON design that
might be improvements for a muon collider are
\begin{itemize}
   \item  The muon collider needs both charges
of muons, while protons produce predominantly
$\mu ^+$ (from $\pi ^+$). This could be solved
by using deuterium ions instead of protons.
   \item  There is no need to be above the
energy threshold for kaon production, and
nucleon (proton or neutron) kinetic energies
as low as 700 MeV produce pions
copiously\cite{pi prod}.
This would be cheaper, would decrease the
decay length of the pions and would decrease
the energy flux onto the production target.
It would also open up the speculative possibility
of using an induction linac instead of a storage
ring for accelerating the protons/deuterium ions.
(Induction linacs can produce accelerating gradients
in excess of 1 MeV/m and reach good efficiencies of
better than 50\% for short, intense particle
bunches\cite{induction linac} -- which sounds
ideal for a muon collider.)
\end{itemize}

    The thermal shock on the target is a difficult
design problem. A bunch of $10^{14}$ 1 GeV protons
delivers 6000 joules onto the target spot in
a nanosecond timescale, some fraction of which
will go into shock heating of the target.
This load is repeated 10 times or more every
second. This must be handled by maintaining
a large spot size and intensive cooling of
the target. A more exotic option which has
already been tested at accelerators is using
a liquid jet target of either water or a
molten metal.

\begin{figure}
   \vspace{10 cm}
   \caption
{A schematic diagram of the beam-line elements used for
pion collection and decay to muons.}
   \label{fig:decay channel}
\end{figure}

    A schematic diagram of the pion collection
and decay channel is shown in figure~\ref{fig:decay channel}.
One speculative alternative is to use a long
($\sim 50-100$ m) solenoidal magnet with a large aperture.
The transverse momenta of the pions coming off the
production target range up to around 300 MeV/c.
Almost all of these pions would be confined
in spiral orbits by an iron solenoidal magnet
with a 2 Tesla field and 50 cm aperture radius,
or by a superconducting magnet with a 6 Tesla
field and a 20 cm aperture radius. The pions would
decay to muons inside the magnet, and the
positive and negative muons could be separated
by including an additional transverse magnetic
field. This idea would be much more practical
if r.f. acceleration could be provided inside
the magnet (I have no idea whether this is
possible). In this case the acceptance could
be a large fraction of unity for both $\mu ^+$
and $\mu ^-$.

    The acceleration of the muons must proceed
relatively quickly to avoid losing too big a
fraction to decays. The average accelerating    
gradient required is several MeV/m, which is
easily within today's technology since the
SLC electron linac currently operates with
an average gradient of 20 MeV/m. A simple
numerical integration finds that when
muons are accelerated from 300 MeV to 2 TeV
at a constant gradient of 5 (or 10, or 20)
MeV/m the fraction surviving is 74\% (or 85\%, or 93\%).

\begin{figure}
   \vspace{20 cm}
   \caption
{Conceptual diagram of a recirculating linac accelerator
structure.}
   \label{fig:recirc linac}
\end{figure}

    Figure~\ref{fig:Neuffer acc} uses a linac
to accelerate the muon beams. This is likely
to be a very expensive option -- almost half
the cost of a $\ee$ linear collider just for
acceleration. Bob Palmer\cite{Palmer talk} suggests
using instead a recirculation linac, as shown
in figure~\ref{fig:recirc linac}. The particles
pass through each of the superconducting linacs
several times over, and
are transported between the linacs by the
bending magnets in the recirculation loops.
The motivation for this design is that r.f.
accelerating cavities are very expensive, so
it is cheaper to use the same cavities several
times per bunch. This design is basically a
higher energy copy of the existing CEBAF
$\ee$ accelerator, which also uses
superconducting r.f. cavities.

   After acceleration the $\mu ^+$ and $\mu ^-$
bunches are injected into the collider rings
in opposing directions. Since muons are heavy
enough that synchrotron radiation is not a
problem their beam transport properties are
similar to protons. For example, 1 TeV muons
would require a ring of radius about 1 km,
being the same energy as the protons in the
Fermilab Tevatron accelerator. The decay
length of the muons in the ring is given by
\begin{equation}
{\rm decay length\: =\: 6233\: km\: \cdot\: E_{\mu}\, (TeV)}.
\end{equation}
This means that the number of muons in a
bunch decays by a factor of 1/e in about
1000 turns -- independent of energy.

   One advantage for muon colliders
over hadron colliders is that the storage time
required is only milliseconds rather than
hours, so the requirements on beam stability
are much less demanding. Palmer suggests using
an ``isochronous'' ring, with few r.f.
cavities to compress the bunch length.

%

\section{Detector Design Issues}

    The particle detectors at the interaction point
would be expected to be similar to those at other
high energy colliders, with particle tracking in
a magnetized space surrounding the interaction
point and with calorimeters enclosing this region.
(One difference might be a greater emphasis on
the precise determination of muon momenta.)

     The backgrounds emanating from the vertex
itself would be expected to much smaller than
for hadron colliders, and probably smaller than
at TeV energy electon colliders. However,
the decay of the muons to electrons will still
lead to serious backgrounds at the detectors.
For 2 TeV muons approximately one in $10^7$
will decay per meter, so a bunch of $10^12$
muons will produce about $10^5$ electrons per meter
with an average energy of about 2/3 TeV. All of
these electrons will eventually hit the beam pipe
somewhere in the ring, initiating electromagnetic
showers.  This leads to two types of backgrounds
\begin{enumerate}
   \item  The electromagnetic showers from electrons
striking the final focus magnets close to the interaction
point can leak into the detector.
   \item  Electromagnetic showers anywhere along the
straight sections before the interaction point
will occasionally produce a muon pair.
This is suppressed relative to $\ee$ pair production
by a factor of $(\mmu / \me )^2 = 40,000$, but the
muons can pass through any shielding placed in front
of the detector.
\end{enumerate}

     These backgrounds must be suppressed by a
combination of shielding and design of the final
focus magnets, and the detector must have enough
electronic channels of tracking and calorimetry
to be able to correct for the remaining background.

     A reasonable design for the
beam-line\cite{Palmer comm.}
might include a final focus region
consisting of iron quadrupole magnets
many meters long with a conical aperture decreasing
from several cm at the entrance to about 1 mm at the
end closest to the interaction point. Much of the
remaining 1--2 meters distance to the interaction
point might have a small aperture
surrounded by a tungsten shield. The thickness
of the tungsten would be determined by a compromise
between the background suppression and the loss
of angular acceptance into the detector. Such
tungsten shields have also been discussed for TeV
scale $\ee$ colliders, blocking up to 10 degrees
of angular acceptance about the beam-pipe.
%
%
\section{Spin-off Physics Opportunities at a Muon Collider Facility}

   A muon collider facility would provide for much
useful physics research apart from muon collisions.
Further physics topics include
\begin{itemize}
   \item spallation neutron experiments
   \item neutrino physics
   \item muon fixed target physics.
\end{itemize}

   The short intense bunches of deuterium ions
used for creating the pions are also ideal for
producing neutrons, and designs for spallation
neutron sources include just such a
beam\cite{Bauer talk}. The neutrons could
either be collected from the primary proton
target or from the beam dump downstream of
the target.
Neutrons are somewhat complementary to x-rays as
important probes for condensed matter experiments,
and the interest in neutron sources is illustrated by
the plans to build the Advanced Neutron Source in the
U.S.A. at a cost of over 1 billion dollars.

    Muon decays in the accelerator straight sections 
around the interaction points would provide a neutrino
source unique in its intensity and composition.
Each cycle of the muon bunch would produce
sub-nanosecond bursts of roughly $10^7$
$\numu$'s and $\nuebar$'s
(or $\numubar$'s and $\nue$'s for the $\mu ^+$
bunch traveling in the opposite direction).
These would have an average energy of around
1/3 the muon beam energy, and would have an
angular divergence of only about
$1/\gamma _{\mu} \sim 0.1$ mr or the angular
spread in the muon directions along the straight
section (whichever is larger). This would
allow substantial improvements in both
precise measurements and seaches for
exotic physics processes in neutrino-nucleon
scattering. For example,
the large neutrino-induced event
samples could substantially improve current
measurements of nucleon structure functions
and weak mixing angle measurements from
neutrino-nucleon scattering, and the purity
of the beam and the 50\% component of electron
neutrinos would allow unprecedented sensitivities
in detector-based searches for neutrino
oscillations (a topic which is currently
popular). In fact, the neutrino beam would
be strong enough to be a radiation hazard,
and it is likely that human habitation would
have to be forbidden along a line extending
out from the accelerator straight sections.

%
%
%
\section{Feasibility and Cost}

\begin{table}[h,t,b]
\centering
\begin{tabular}{|l|c|c|}
\hline
 parameter                         &   muon I           &   muon II   \\
\hline
luminosity  (${\rm cm^{-2}s^{-1}}$)& $1.3 \times 10^{33}$ & $4 \times 10^{34}$\\
beam energy  (TeV)                 & 2                  &  2  \\       
proton frequency (Hz)              & 10                 &  30 \\
protons/bunch                      & $6 \times 10^{13}$  & $2 \times 10^{14}$ \\
muons/bunch                        & $4 \times 10^{11}$  & $1 \times 10^{12}$ \\
phase space (${\rm MeV^3\, mm^3}$) & $1.0 \times 10^5$   & $0.8 \times 10^5$  \\
\hline
\end{tabular}
\caption[]
{Parameter choices for a muon collider~\cite{Palmer talk}. }
\label{para table}
\end{table}

    The parameters of two conceptual designs for a muon
collider by Palmer\cite{Palmer talk} are given in table~\ref{para table}.
Achieving the design luminosities given
by Palmer would make such muon colliders extremely
attractive for exploring the TeV energy scale.
It should be stressed that a lot of work
will be required before one can estimate with
any confidence what are reasonable design
parameters for a muon collider.

     Palmer also provided an ``order of magnitude''
cost estimate for a 4 TeV CoM muon collider, with
the caveat that it was an extremely crude estimate
which should not be taken seriously. He obtained
the proton source cost (0.5 billion) using the KAON
cost estimates, the linac cost (1.0 billion) using
estimates for the Next Linear Collider $\ee$ machine
and the tunnel and magnet cost (0.2 billion + 0.9
billion) by scaling to the SSC.
Adding 0.5 billion dollars for the facility and
0.3 billion for the muon cooling gives a very
tentative estimate for a total  cost of 3.4
billion dollars. This is certainly a very hefty
price tag, but it is competitive with and probably
cheaper than the competing technologies, and the
price would be less for a lower CoM energy.

\section{Summary}

    Muon colliders show great promise for exploring
the the high energy frontier in elementary particle
physics. However, it will take a lot of detailed
study to determine whether they are actually feasible
or are just another good idea that won't quite work.

%

\pagebreak

\bibliographystyle{unsrt}

\end{document}